\documentclass[12pt,a4paper]{memoir}
\usepackage[T1,T2A]{fontenc}
\usepackage[utf8]{inputenc}
\usepackage[greek.ancient,latin,german,french,dutch,italian,spanish,russian,english]{babel}

\DisemulatePackage{setspace}
\usepackage[onehalfspacing]{setspace}

\usepackage{filecontents}

\usepackage[backend=biber,citestyle=verbose-trad1,bibstyle=verbose-trad1,language=auto,autolang=hyphen,autocite=footnote,citetracker=true,ibidtracker=false,idemtracker=false,opcittracker=false,citereset=chapter,citepages=omit,isbn=false,url=false,doi=false,eprint=false,ibidpage=true]{biblatex}
\usepackage[style=english]{csquotes}
\usepackage{calc}
\usepackage{textcomp}
\usepackage{hyphenat}
\usepackage{microtype}
\usepackage{keyval}
\usepackage{ifthen}
\usepackage{url}
\usepackage{etoolbox}
\usepackage{nicefrac}
\usepackage{endnotes}

\emergencystretch=\maxdimen
\AtEveryBibitem{\clearfield{note}}
\AtEveryCitekey{\clearfield{note}}
\AtEveryBibitem{\clearfield{isbn}}
\AtEveryCitekey{\clearfield{isbn}}
\AtEveryBibitem{\clearfield{url}}
\AtEveryCitekey{\clearfield{url}}
\AtEveryBibitem{\clearfield{lccn}}
\AtEveryCitekey{\clearfield{lccn}}
\AtEveryBibitem{\clearfield{doi}}
\AtEveryCitekey{\clearfield{doi}}
\AtEveryBibitem{\clearfield{eprint}}
\AtEveryCitekey{\clearfield{eprint}}
\AtEveryBibitem{\clearfield{issn}}
\AtEveryCitekey{\clearfield{issn}}
\AtEveryBibitem{\clearlist{publisher}}
\AtEveryCitekey{\clearlist{publisher}}
\AtEveryBibitem{\clearlist{language}}
\AtEveryCitekey{\clearlist{language}}
\AtEveryBibitem{\clearfield{note}}
\AtEveryCitekey{\clearfield{note}}
\usepackage{hyphenat}
\setsecnumdepth{section}
\setsecheadstyle{\large\textbf\centering}
\setsubsecheadstyle{\large\centering}
\usepackage[hang]{footmisc}
\usepackage[twoside=false,bindingoffset=1cm,left=2cm,right=1cm,top=2.5cm,bottom=2cm]{geometry}
\usepackage{calc}
\usepackage{textcomp}
\usepackage{microtype}

\DefineBibliographyStrings{english}{
	pages = {p\adddot}}
\DefineBibliographyStrings{french}{%
	pages = {p\adddot}}

\usepackage{stringenc}
\usepackage{pdfescape}

\DeclareNameAlias{sortname}{last-first}
\DeclareNameAlias{default}{last-first}

\author{Starostin D. N.}

\title{Astronomy of the Earth-Moon system and the Eschatological Expectations of the Christian Historians of the 5th century CE}

\date{}

\begin{document}

\maketitle

\begin{abstract}

This is a historian's view of how modern astronomy data can be used to discuss the shifting historical worldview of Late Antiquity. In this article an attemp is made to construct an approximate model of how the cycles of astronomical bodies' visible rotation aaffected the writing of history and self-representation of the Roman empire's powerful people. It is argued that while rare outstanding events like solar eclipses might have caused a short stir in the minds of the rulers and their environment, long-term cycles based on the synchronization of the Moon's phases with the solar calendar and the cycles of the planets lining up in the same disposition (in relationship to the Moon or without this relationship) were the foundation of astronomy-based Christian chronological system. The emergence of the Christian historical worldview in the 5th century was marked by appearance of a significant eschatological strain in it. Historians paid attention not only to the theology-defined signs of the end of the world, but also, as it has been suggested in modern studies, to the some outstanding celestial phenomena. In this paper I would like to address several criteria which may help understand what in the celestial motions interested the astronomers and historians of the 5th century. This paper uses the first approximation of astronomical data for solving the problem of how relevant the skies were for historians, although all numeric parameters are taken from the up-to-date astronomy reference publications. It is an attempt to understand whether the very basic approximations can be related to what historians know from the array of sources available to them. The analysis suggests that there is a whole array of occasions when the dates of astronomical events, received with the help of these basic approximations, coincide with the data from historical sources.

\end{abstract}

In Late Antiquity, Christian history emerged from the traditional to the Classical Greek and Roman world historiography. Historians started to reconcile biblical chronology with their own since the the time of Hellenism.\autocites[451-452]{Wacholder:1968} Julius Sextus Africanus made a significant effort in the 3rd century to produce an uniform chronology of the Old Testament and of the Christian era.\autocites{Africanus:2007}{Gelzer:1898}{Mosshammer:2006}{Mosshammer:2008} The writing of Eusebius, Ammianus Marcellinus, Orosius, Sozomenos, Socrates, Philostorgios, Hydatius, Sulpicius Severus, and Prosper of Aquitaine created a foundation in the form of a chronicle.\autocites{Brooks:2014,Burgess:1999,Burgess:2011,Burgess:2012,Burgess:2012a,BurgessKulikowski:2013,Croke:1983,Croke:1987,Croke:2005,Zecchini:2003} Theologians like Augustine and the historians who followed in his path (Orosius, Hydatius, Sulpicius Severus, Prosper of Aquitaine) contributed to write a new historical narrative that superimposed the Christian vision of history, with its resonant eschatological theme, on the histories of the Roman empire and the regions comprising it. The presence of eschatology in historical thinking was addressed by Richard Landes, who argued that it was indeed an important cultural and religious paradigm for the Christians since the 5th century.\autocites{Landes:1992,Landes:1993,Landes:2000}

In a recent study it was suggested that the astronomical phenomena took a significant part in the narrative structure of the 5th-century historians. Thus Hydatius’ observing the blood Moon of 462 CE (a usual phenomenon when the Moon is its perigee) was placed in the context of the barbarian ruler, Euric, grabbing the power in Spain in 467 CE and the failed naval expeditions of Leo and Anthemius in 468 CE.\autocites[21]{Wieser:2018} Thus this Iberian historian successfully put the eschatology as an abstract context to the test of the astronomical observation. It has also long been noticed that in the 8th century Venerable Bede paid attention to the fact that during one of the eclipses of the 7th century the Moon reached its phase in discord with the Easter tables.\autocites{Moreton:1998}

In the light of the new advances in the astronomy I would attempt to draw attention to one particular coincidence that may theoretically serve as an explanation of the 5th-century’ particular disquietude in the writing of history. Many studies have laid the theoretical groundwork for solving the problem of the Earth-Moon system in precise mathematical terms.\autocites{Euler:1750, Euler:E187:1753,Hansen:1838,Newcomb:1898,Brown:1896,Brown:1926} An analysis of these theories was summarized by F.~Tisserand.\autocites{Tisserand:3:1894} By 1984 modern measurements and calculation methods removed all empirical terms from the Moon’s ephemeris\autocites{Meeus:Savoie:1992,Meeus:2009, Morrison:Stephenson:2004,Simon:1994}[317]{Seidelmann:1992} The information gained with the help of modern methods has now been put up for academic use at various sites including the NASA eclipse site (https://eclipse.gsfc.nasa.gov/). This allows a historian to investigate whether history and the astronomical calculations can be both employed to make judgment about the chronology of Late Antiquity. The approach employed in this paper is a first approximation, based on the modern data.

The question that arises when a historian looks at the astronomical information for the period is a dichotomy of what was more important for the astronomers in Late Antiquity, the rare and singular, but spectacular events like solar eclipses or the periodic events like regular lunar phases at expected periods of time and planetary line-ups. It is a long-standing question whether scholars in Antiquity managed to find a way to include solar and lunar eclipses into a pattern of a calendar. In other words, it is a question of whether people of Late Antiquity chose to be scared of the rare celestial events and to interpret them as divine punishment or they favored stability and made a system of continuing and repeating events. I argue that the patterns of lunar, planetary, and stars' motion played a much larger role than the eclipses. In other words, the periodic motions that were possible to predict with reasonable accuracy played a much more significant role in the relationship of people to the astronomical events than did the eclipses. The basis of calendars were the repeating motions of the celestial objects.

Let us first consider the arguments about the importance of eclipses for astronomers' and rulers' views on history.

Eclipses, the events that stood out of the calendric patterns in Late Antiquity, did not seem to be associated with the Modern scholars have long known two major solar eclipses in the reign of Nero on April 30, 59 CE and in the reign of Constance II in 346 CE (we have shown why the latter was important earlier). One may wonder whether such natural phenomena could have influenced the underlying stability of human perception of time. In regards to the eclipse of Nero’s time, Tacitus mentioned that even the Sun chose to hide from the human sight in light of the emperor’s killing of his mother (Tac. 14:12; Pliny Hist. Nat. 2:70).\autocites{Lynn:1893}[\nopp 14:12]{Tacitus:Annals}[\nopp 2.70]{Pliny:NH} But there was nothing of this kind in 346 CE when the eclipse was total for at least one area of the Roman empire, Antioch.\autocites[3-5]{Hayakawa:2022}

The examples of astronomical information one can find in the works of Late Antique chroniclers and their continuators suggest that the facts of eclipses became part of the historical narrative tradition and even started to have some eschatological meaning by the end of the 4th century and that historians learned how to describe their influence on the populace and troops correctly. But although they took heed to celestial phenomena, they left the attempts to find a mathematical rule to later generations. Jerome included an eclipse in his chronicle without a correct date, so the possible candidates are June 6, 346 CE or October 9, 348 CE. In the 19th century scholars thought the eclipse of 346 CE to have been visible throughout the Roman empire, while recent studies suggest its totality spot was in Antioch.\autocites{Lynn:1893}[3-5]{Hayakawa:2022} But according to modern calculations there must have been several eclipses in the vicinity of this date, which must have been visible to at least some residents of the Roman empire: these were the eclipses of the years 341, 344, 345, 346, 348, 349, 351, 354, 355, 356, 358, 360.\autocites{Espenak:Meeus} The 346 CE eclipse reached the phase of totality only in Antioch, as scholars think now, and other eclipses may have been even less visible in Constantinople. The eclipse of 346 is cited in the early 9th-century chronicle of Theophanes the Confessor, but it is a much later text that must have had its antecedents in regards to this information. Some scholars believe that the original text that mentioned the eclipse was written by Eusebius of Caesaria’s pupil Eusebius of Emesa (ca. 330–ca. 360 CE).\autocites{Reidy:2015}[273]{Burgess:1999} This 4th-century scholar was a court astrologer for emperor Constance II, in addition to being considered one of the important theologians of his period. Since he was accused of sorcery, it may well be that it was during his tenure as the court astrologer that several partial and annular solar eclipses took place (Wynn, 2011). In other words, the information about the 346 CE eclipse (or several eclipses), which might have had eschatological underpinning did not enter the historians’ parlance and became the matter of historical narrative only well after the reign of Constantius II had ended and after the Roman empire in the West had ceased to exist.

In contrast, Ammianus Marcellinus more precisely pinned down the eclipse in the year 359 or 360 CE in his “Res gestae” 20.3.1-2 and, unlike Jerome, spoke of the nearly eschatological outcome of this event for the Roman army.\autocites{Ammianus:1978} Some scholars believe his description to be impeccable from the point of view of Late Antique astronomy, while others argue that Ammianus was not quite correct in discussing the difference between the New Moon and the eclipse.\autocites[20.3]{Szidat:1977}[141]{DenHengst:1986} If we are to believe the dating and his information, it must have been the eclipse of August 28th, 360. But since Ammianus spoke of how Julian rallied a large number of Gallic legions to move to take part in the campaign in the East, the August date seems a bit late. Is it possible that Ammianus confused this eclipse with the hybrid eclipse on March 15th, 359, a year before that? This eclipse (a hybrid one) may better conform to the description of the historian who spoke of how the Sun’s surface was “cut off by a spear” (lancea). In this case the events hat had been brought into motion could have lasted from March 15, 359 to 360, and it was enough time for Julian to understand tensions among soldiers and procurement problems and to make a decision to propose them to go on a campaign in the East with their wives. The requirement to arrange for public postal carts for them to relocate also required a significant amount of time. In other words, it seems that in Ammianus Marcellinus’ narrative the eclipse took a significant place as an event that caused some stir in Gaul. Thus unlike Jerome, who shunned from giving any social commentary on what the eclipse might have caused in the population, by the end of the 4th century Ammianus could venture on describing what a celestial event meant for the Roman army in the West, where the eclipse was the most pronounced. This historian, in contrast to Jerome, already became well-versed in the writings of Ancient and contemporary astronomers as to explain the reasons for an eclipse in a clear Latin\autocites[20.3]{Szidat:1977} Ammianus may have made even more use of it because the eclipse served as a kind of forewarning of Julian becoming a usurper. It seems that the authorities’ and community’s response to such events only became sensible if there was a consensus among the the powerful people and scholars on how to react to celestial events. Ammianus’ approach to this event suggests that this consensus was becoming crystallized, as an out-of-order celestial event became a commentary on the outsized ambition of formerly modest Caesar.\autocites[21.5.1]{Szidat:1977} Other remarkable eclipses of 418 CE (described by Philostorgius) and of 484 CE show that historians learned how to make sense out of celestial events, but also managed to stay clear of the overarching conclusions with eschatological implications.\autocites[5-7]{Hayakawa:2022} In other words, by the early 5tn century eclipses and the conjunctions of the lunar calendar with the solar one became a phenomenon historians could include in their narratives and even make it a sign of the Christian history approaching a predetermined stage.

One may construct an argument that astronomers sought how to accommodate the eclipses into the repeating patterns of time. But the task of predicting them was largely impossible. Instead, by the end of the 4th century historians managed to construct a language that would help them describe solar eclipses as both natural and social phenomenon, the latter having strong eschatological notes. The eschatological aspect of these commentaries betrayed in the first place a desire either to fit an event into an already functioning set of cycles as a real marker of a repeating phenomenon or to discard it as a meaningless fluke.

In other words, the introduction of Christianity as the official religion and of its lunisolar calendar may have been responsible for making scholars pay more attention to the correlation of the Sun and the Moon’s cycles and their calendars. This did not happen because they forewarned the “judgment of Heavens”, but for another reason, I believe. What was more important were the repeating patterns in the motions of the Sun and the Moon together, or rather, the situations when the lunar phases fell on the same days of the lunar calendar as it did at some historically or symbolically important date like the foundation of the Temple by David or the birth of Christ. The specialists in time reckoning knew about the repeating patterns of 19 and 76 years. Solar and lunar eclipses fell out of this pattern and they were hard to predict due to their irregularity at a given location. In other words, I argue that the solar eclipses that were hard to predict were not intepreted as part of a picture of universal history, on which a community's self-representation was built. But the repeating patterns of the conjunctions of the solar and and lunar calendars did indeed attract more attention than solar eclipses from astronomers and historians, who seemed to attempt to aling the political events with the conjunctions of the solar and lunar calendar.

Because the astronomers of Late Antiquity may not have known that the actual length of the year was 365.24219 days and that the lunar month was actually 29.53 solar days (29½ days plus ca. 43 minutes), there accumulated in calendars a small but significant discrepancy. If the discrepancy was not counted in, the solar calendar lost one day over the course of about 276 years. This seemed to be the problem of the Egyptian and of the Julian calendars, which held the length of the year to be exactly 365 and a quarter day. The same happened to the calendar of the Moon phases and the projected calculations must have been routinely off the actual phase of the Moon at the expected time on the given day of the solar calendar.\autocites[398]{Locher:2006} Since the precession data is incorporated into most modern calculations,\autocites{Precessions:1976} we will not be using it directly. For the purposes of advancing our hypothesis we will using the modern mean values. In the long run, as we will show, the mean values provide a fairly good approximation to explain how the celestial bodies were visible to astronomers and people in Late Antiquity. Let us use a simplest model here without employing the $\Delta$T calculations. These will be at first purely hypothetical considerations which we will later compare with the data available to us from historical sources. This approach is justified because it is from the historical sources that modern astronomers take the first information to construct their models.

The mathematical formula for that is (‘i’ is the integer number of years): $|(i \times 365.24219) mod 29.53)| \leq 0.25 or |(i \times 365.24219) mod  29.53)| \geq 29.27$, which means a conjunction of the synodic positions of the Sun and the Moon within ¼ of a day, that is, within 6 hours. The case in point here are the calculations that would be made from the year 1 BCE as the measure of the Moon’s phases across the Julian calendar’s year roll. If we take the actual astronomical solar year and the lunar month, then, without the imprecision that the human-made calendar introduced, there were several dates when the solar and the lunar rotations coincided. These were the years 19 (the foundation of the Easter calendar), 38 (two 19-year cycles), 57, 76, 483, 502, 521, 540, 559, 578 and 598. Let us notice that there was a long hiatus of 3 centuries when the phase of the Moon did not repeat its showing on 1 BCE in regards to the same dates of the Solar, Julian (or in some cases, proleptic Julian) calendar. Let us also notice that several dates fall in the late 5th and the early 6th century, the period that was marked by the Fall of the Roman Empire in the West, the re-fashioning and consolidation of the Byzantine empire in the East and by the creation in the West of several barbarian kingdoms. Frankish king Clovis I came to power in 481 CE and was gradually gaining his prestige, ultimately defeating Syagrius in 486. The year 483 was also important because it was close to the cycle of precession of 476 years that L. Euler matematically calculated. It was alro the year when the Easter calendar had to be adjusted to take one day off the count of lunar epacts.\autocites[400, \S 379]{Euler:1772:E418} The year 521 CE was the time when early medieval scholars started thinking about the new Easter calendar that was later confirmed and realized by Dionysius Exiguus. 540 CE witnessed the peak of Byzantine military campains of Belisarius, ordered by Justinian.

In addition to that, there was a discrepancy between the astronomical calendar and the Julian calendar, which could become expressed in days. But in some cases the problem was less in the actual count of days since in the Ancient world the astronomers managed to create calendars that helped intercalate any extra days into the year (with an embolistic month and a saltus lunae).  The problem also lay in the extra hours that cumulatively added up from the minutes and seconds of the discrepancies between the Julian solar calendar and the Ancient world’s lunar calendar with the actual astronomical phenomena. These discrepancies between the astronomical calendar and the Julian calendar can be calculated at the first iteration with the help of the modern data for the parameters of the Earth’s and Moon’ orbits. The simplest mathematical formula that is used here is of the following form: $|(i \times 365.24219) mod  29.53) - ((i \times 365.25) mod 29.5)|$. Its use is justified as the first iteration approach to the problem and is supported by works on astronomy from Leonard Euler on. The actual astronomical conjunction of the Moon’s phase with the solar synodic position in this case came on approximately the following years: 77, 156, 233, 312, 390, 467, 544, 623, 700, 778. Thus one may notice two separate sequences: one of purely astronomical conjunctions, some cycles of which govern the lunisolar calendar today, and the other of the correlations (with half a day precision) between the astronomical and the Julian calendar. These sequences do not count in the longer precessions of the Earth’s and Moon’s rotation with the length of 1000 years and so on, but L. Euler’s works on the Earth-Moon motion showed that these periods do exist in the heuristic observations and calculations with all types of precession included as a parameter. These years are themselves an approximation as normally there were several years in the vicinity of each of these cardinal dates when the Moon’s phases were within one day of those in the year 1 BCE as measured against the Julian calendar. It can be argued that the most common average period when both the astronomical conjunction and the correlations  between the astronomical and Julian calendars took place was between 72 and 84 years, with the most weight of the phenomenon falling on 76 years.

The question of employing the conjunction of the solar and lunar calendars for the better observation of the Moon’s phase seemed to have bothered historians and computists since the 4th century. Of the dates of the conjunction, two (313 CE and 390 CE) are quite close to the two most egregious cases in Jerome and Prosper’s chronicles where they either hinted at or showed a one-year discrepancy with other sources. Eusebius put the battle of the Milvian bridge at the 7th year of Constantine, which was 312 CE, while Jerome set it to the 6th year of Constantine, although it was also 312 CE, since he had set the 1st year of Constantine at 307 CE.\autocites[43]{Burgess:1999} In this case the Moon’s calendar for 312 CE and for 313 CE was within one-day’s precision in regards to the calculations from the year 1 BCE, but the difference was, respectively, 4 and 13 hours. So Jerome’s indecision to choose between the 2 dates is understandable as he sought to move the start date of the count so as to make the discrepancy less. Exactly 6 years had a discrepancy of only 6 hours with the year 1 of Constantine, while the year 7 had a much larger discrepancy of 15 hours. So for an important event like the battle of the Milvian bridge he needed that it be at the beginning of the year 6 of Constantine so that the position of the Moon be predictable in the the terms of the conjunction between the solar and the lunar calendars. L. Euler in the 18th century determined one cycle of the Moon’s motion to be 6.5 years, which explains that there was a conjunction of the Moon’s position in the middle of the 7th year, but before its end.

Prosper of Aquitaine, in turn, moved the events from 383 CE to 384 CE.\autocites[166]{Humphries:1996} The shifting of an event from 383 CE to 384 CE was meant better to account for the fact that in terms of the lunar phases, the Solar calendar was running out of sync in 383 CE and it came in sync by 384 CE. In his case the reason might have been that in the year 383 CE the discrepancy was 9 hours, while in the year 384 CE the Moon reached near total conjunction with its position in the 1 BCE in terms of hours, while the visible lunar epact had grown by 2 days. In other words, the Full Moon could be seen at the same time at night as it was in 1 BCE. This was the case for the years 389 CE and 390 CE, too, where the discrepancy between the calculated position of the Moon and its real phase was minimal (plus 3 and minus 5 hours respectively). It is quite likely that full days did not bother computists at all since there were mathematical means to account for them. Thus one may notice that since the 4th century, when Christianity and its lunisolar calendar became official, historians did care about using the dates for which the projected correlation between the motions of the Moon and the Sun was in nearly total sync with their visible correlation. In other words, the first two centuries of the Christian Roman empire raised a number of significant challenges for historians and computists because they had to address the problem of making a time to observe the Moon at the time when its phase coincided with the expectations made of the basis of the “Egyptian” calendar. In discussing the problems of the Christian time reckoning in the 4th and 5th century one may consider the discrepancy that emerged because of the actual astronomical calendar having a shorter year than the one used in the Egyptian and Julian calendars.

In other words, I argue that the innate 76-year cycle in the Earth-Moon system’s motion and the patterns of the Moon's phases exactly coinciding with the dates in the Solar calendar (like the Vernal equinox) and thus taking the same relative position in the sky against the immovable objects on the ground (temples) made astronomers and historians look after the solar eclipses with increased attention. It was not vice versa.

One may also consider another factor in discussing how the celestial events might have influenced the vision of history among the educated people. In addition to the conjunctions of the solar and the lunar motions one may also consider the lineup of planets in the same form as a possible marker of information the astronomers gathered by looking at the skies. Although the lineups of planets are a common event, the cases in which it took place on the same date of the solar calendar are much more rare. The conjunction with the solar calendar is important because it is connected to the Vernal equinox, which had been observed for the purpose of relating the solar and the lunar calendar to each other. For the purposes of this paper I will only consider cases when the lineup of planets was observed during a solar eclipse, since there is very little information on what astronomers saw in the night skies on usual days. Thus we need to consider the eclipses of the Ancient world as the possible starting points of observation when the astronomers remembered he configuration and made it a historical point of reference. These cases are visible with the help of modern astronomical tables that use the advanced mathematics developed since the 17th century.

There is another reason for the years in the vicinity of 476 CE (like 468 CE and 483 CE) being of interest to astronomers and historians of Late Antiquity. In terms of the astronomical observations the 5th century was special because during it the long-term cycles of the Earth's and other planets' motions had come to a synchronization. The rotation of the planets around the Sun makes them line up in various configurations in the skies visible from Earth. Already in the 13th century BCE the Egyptians were watching the planets.\autocites[144]{Krauss:2002}[378]{Krauss:2006} The period in which a planet returns to the same point in the skies as visible from Earth (the synodic period) has now been precisely calculated by astronomers. The synodic periods for Earth, Jupiter and Saturn are very close, while the periods for Venus and Mars differ significantly. As the first approximation of solving this problem is the approach which seeks to establish the level of synchronicity of the planets' synodic periods by examining their periodic qualities against that of the Earth's orbit and of the solar year. But we are interested not in all the lineups, which happen regularly about every 20 years, but only in those that happen at exactly the same date as the original one (say, the New Year, the Vernal Equinox, 1st September, or any other civic holiday). If one calculates the offsets, i. e. the discrepancies between the length of the solar year and synodic period for each planet, this data can be used to determine when the planets would reappear in the same order and at the same visual distance from each other as they had once been observed. Naturally, those years when the offset is the smallest are the years when the configuration of the planets was the same. Setting the offset to within 3 to 4 months is a reasonable assumption. There is a further option to take the offsets, calculate the geometric mean and to calculate the mean square deviation for each of the planets' offsets. Finding the years when this mean square deviation is less than a month is a good technique to find those years when the planets were in the same line-up as initially. Interestingly, both methods give approximately the same results. For the purposes of this paper the possible precession of the planets’ synodic position due the Earth’s precession against the fixed stars is not considered since it is the first approximation and since the proposed method does give heuristically verifiable results within the given limits of approximation.

Calculations would not mean much if we would not set a starting point. Solar eclipses in Antiquity are a good starting point since during an eclipse the line-up of planets became visible during the day and attracted attention of the educated people and the populace. Once remembered, this line-up could have long stayed in memory. The ancient documented solar eclipse of 2161 BCE (Huber, 1999), that of Shamshi-Adad of 1587 BCE (probably the eclipse of Oct. 23, 1588 BCE or of Mar. 28th, 1586 BCE)\autocites{deJong:2013}, the recorded eclipse of Sep. 11, 1557 BCE\autocites{Henriksson:2005}, the lunar eclipse of 1547 BCE (the Ur III eclipse)\autocites{Banjevic:2006}{Banjevic:2012}, the lesser-known solar eclipse of Feb. 15, 1547 BCE\autocites[3]{Eclipse:Babylon:2020}, the solar eclipses of Oct. 30th, 1207 BCE (Joshua 10:12-13)\autocites{Vainstub:2020}, 1096 BCE and 1084 BCE (March 27th) may be taken as possible candidates. To understand which of the starting points is better one needs to use modern data. For example, since 2023 is the year of the planets' line-up, it is remarkable that this configuration of planets repeats the order that might have been visible during the eclipse of the year 2161 BCE or of 1084 BCE, but not during the other ones. For the purposes of this paper we obviously do not count in any planetary precession since these considerations are the first approximation.

Calculations which involve finding those years that have the smallest offsets in terms of the planets' synodic periods first show the importance of years in the vicinity of 483 CE. For example, if calculated from the solar eclipse of 1547 BCE, in 484 CE Jupiter came earlier by 18 days, Venus and Mars appeared right where they had been in 1547 BCE, while Saturn was early by 160 days. Let us notice that the Solar eclipse of 484 CE was well observed (Hayakawa et al., 2022, 5–7). If counted from 1084 BCE, the planetary configuration repeated in 483 CE and ca. 601 CE, when another eclipse was visible in the Mediterranean.\autocites[7-8]{Hayakawa:2022} If calculated from the Solar eclipse of 27th March 1084 BCE, in the year 483 CE Jupiter and Mars both came to the same position in the sky 116 days ahead, while Venus was ahead by 70 days. In other words, in the period around Christmas those interested in heavens could observe the planetary line-up that was around Vernal equinox ca. March 27th, 1084 BCE. This is the result we get if we calculate the offset's absolute value against the length of the solar year. Using the mean quadratic deviation against the geometric mean of the offset we get much shorter offsets against that day in the solar year when these planets seemed the closest to their original position in the starting year. 

One of the periods that emerges from calculations of this kind is that of 1107 years. It deserves particular attention because the end of it, if counted from one of the eclipses of the Ancient world, and also from the founding of Rome, fell right into the middle of the 4th century. An eclipse could have made the astronomers notice the same lineup of planets as it was during the past event. Eclipses like that may have also let the astronomers of Late Antiquity to pay attention to the celestial events and to notice a flaw in the civic calendar that made it lose one day over 276 years. A well-known Solar eclipse took place in Nineveh on June 15th, 763 BCE (it was also the day of the New Moon, so the two events coincided to produce a sense of .darkening in the skies. In the next year, 762 BCE (761 astronomical year) a Full Moon was on March 22nd, one day after the Vernal equinox. When a whole sequence of solar eclipses started in 346 CE, the Full Moon was on March 23rd, one day late. These dates would have had no specific importance, if the period of 1107 years would not be remarkable for being one of those periods when the planets showed up in the same order and position as they were initially. It was the 7th year of the cycle if the 19-year cycles were counted from 763 BCE and the 6th, if 762 BCE was considered the beginning of the count. The 19-year cycle, if counted from 762 BCE, started on 340 CE, but there was no near coincidence in that year. In 341 CE (the 1st year of the new 19-year cycle if 762 BCE was the first year) the Full Moon came 4 days earlier than in 762 BCE, on March 19. It was also a year of a solar eclipse\autocites{Espenak:Meeus}[3-5]{Hayakawa:2022} In 346 CE, another year of a solar eclipse and the 6th year of the 19-year cycle counted from 762 BCE, the Full Moon occurred on March 23rd, one day later than in 762 BCE. It is interesting to notice that since the Solar calendar was late 4 days in comparison to the Moon's cycle of rotation in 1103 years (762 BCE to 341 CE) even though this period was an exact integer number of 19-year periods, it was losing 1 day every 276 years, a well-known parameter. In other words, in 1102–1109 years, that is, ca. 341 CE, it became clear that the solar and the lunar calendars in the way they were counted were asynchronous and that the solar calendar lost one day to the lunar calendar in 276 years. But naturally, the Egyptian scholars must have taken some problems out of it during the rule of the Ptolemys' from 305 BCE to 30 BCE and the reform of the calendar under Octavian August must have adjusted all discrepancies. The problem was in the new discrepancy that should have accumulated by the 3rs century. In other words, the period of 1107 years made those interested in astronomy take notice of the discrepancy and of asynchronous character of the calendars. It is all the more important because this visible disagreement in the calendar (4 days) happened ca. 354 CE, which would be the year 1107 from the founding of Rome in 753 BCE.

In both calculations, however, in 483 CE or 484 CE the 4 planets came close to their line-up in either 1084 BCE or 1547 BCE. If the start date was the eclipse of 1557 BCE, then in 476 CE the line-up repeated that at the starting point. Let us notice that one planetary cycle is about 468 years within our simple but rigid criteria (but it is not a repeating one). Thus in the middle of the 5th century the planets lined up as they used to during one of the ancient eclipses, but also in the same fashion as they were at the birth of Christ. 

In the repetition of the planets' conjunctions there are also other heuristically calculable periods of ca. 1054 years, of 1739 years, and of about 3107 years, which give almost complete synchronization within the above-mentioned set of factors. Depending of the start of the calculation and its relation to an eclipse (or lack thereof), the years 476 CE, 483 CE or 484 CE were those of the original line-up that repeated that of 1557, 1084 or 1547 BCE. If the start year was the eclipse of 1207 CE, which the scholars now attribute to that which is mentioned on the Book of Joshua 10:12-13, it gave 3 historically interesting dates of 337 CE (the death of Constantine the Great), 390 CE and 479 CE, as well as the years in the vicinity of the rule of King David. In other words, using the Babylonian and Hebrew chronologies, scholars in the Christian Roman empire might have become aware that in the 5th century there will be one planetary line-up that would exactly repeat that of the beginning of the actual Hebrew chronology that appeared in contact with the Egyptian and Assyrian empires. Even though it was a repeating event, it might have influenced scholars in attracting them more towards observing the skies.
Interestingly, these calculations hint at the intrinsic importance of the year 1108 BCE (minus 1107 astronomical year), when a lunar eclipse happened on December 28th. If the calculations of the planetary configurations were to be applied here, the year 1107 BCE becomes critical because it was when the planets lined up in the same configuration as they did in 2161 BCE because it came 1054 years after the latter (and as it has been stated already, 1054 years make the planets return to the same position within days of each other). Since the eclipse was 3 days away from the start of the new 1107 BCE, the line-up must have been a nearly perfect match and it might have become a start of a new year count. Remarkably, it is claimed to be the last year of the reign of Ramesses X (1111–1107 BCE), which caused a discussion among scholars. The suggestion advanced by a scholar that his reign lasted longer has been rejected.\autocites[493]{Krauss:Warburton:2006} When part of the lunar disk was covered by the Earth's shadow, the astronomers might have noticed one of the planets, probably Venus, that could have been occulted by the bright edge of the Moon's disk. That is, with a degree of precision we may hypothesize that the line-up that people saw in 2161 BCE was also visible for those in late 1108 - early 1107 BCE, and in 405, 457, and 490 CE.

These considerations suggest that Late Antique astronomers and historians built their chronological schemes around long-term repeating patterns in the showing of the Moon and the planets. Solar eclipses that were harder to predict fell out of these patterns and did not seem to influence the chronology or calendar in any way: they were never used as the end of an old era or the beginning of a new one. But when the solar and the lunar calendars repeated themselves and the showing of the two celestial bodies happened in sync (although it was the night for the Moon when the Sun was not visible), this seemed to be the time when scholars and those in power paid more attention to represent themselves according to the history's plan. Thus one may safely argue that for the historians of the 5th century astronomy suggested that the end of times, or at least the end of the long cycles of time, was coming about since the planets and the Moon lined up at the same time and in the same patterns that they did at the time of the constructioh of the temple of David and of the foundation of Rome. It was a factor that determined the interest to the eschatological discourse.

\printbibliography

\end{document}